\documentclass[11pt]{article}
\usepackage{authblk}
\usepackage{acl2015}
\usepackage{times}
\usepackage[utf8]{inputenc}
\usepackage{blindtext}
\usepackage{url}
\usepackage{latexsym}
\usepackage{graphicx}
\usepackage{amsmath}
\usepackage{caption}
\usepackage[colorlinks=true,citecolor=blue,backref=page]{hyperref}
\usepackage{amssymb}
\usepackage{subcaption} %
\usepackage[justification=centering]{caption} %
\usepackage{chngcntr} %
\usepackage{fancyhdr}%
\usepackage[table]{xcolor}  %
\usepackage{multirow}
\usepackage{tabularx} %
\usepackage{tabulary,booktabs} %
\usepackage{ragged2e} %
\usepackage{changepage} %
\usepackage{multicol}
\usepackage{appendix}
\usepackage{hyphenat}
\usepackage{lipsum}
\usepackage{caption}
\usepackage{apacite} %

\title{Multi-Modality in Music: Predicting Emotion in Music 

from High-Level Audio Features and Lyrics}

\author{Tibor Krols}
\author{Yana Nikolova}
\author{Ninell Oldenburg\thanks{\small \hspace{1mm}Authors are listed in alphabetical order and all contributed equally}}

\affil{\small University of Copenhagen, \\\small Department of Nordic Studies and Linguistics \\\small and Department of Computer Science}

\begin{document}
\maketitle
\begin{abstract}
This paper aims to test whether a multi-modal approach for music emotion recognition (MER) performs better than a uni-modal one on high-level song features and lyrics. We use 11 song features retrieved from the Spotify API, combined lyrics features including sentiment, TF-IDF and Anew to predict valence and arousal \shortcite{russell_circumplex_1980} scores on the Deezer Mood Detection Dataset (DMDD) \shortcite{delbouys_music_2018} with 4 different regression models. We find that out of the 11 high-level song features, mainly 5 contribute to the performance, multi-modal features do better than audio alone when predicting valence. We made our code publically available\footnote{\tiny \url{https://github.com/Tibor-Krols/CogSci2-Spotify}}.
\end{abstract}

\section{Introduction}
\pagenumbering{arabic}
The emotional effect of music is a fact both well-studied \shortcite{hunter_music_2010, balkwill_cross-cultural_1999, deng_emotional_2015} and intimately familiar for most people. Furthermore, cross-cultural consistency in emotion recognition in music \shortcite{balkwill_cross-cultural_1999, fritz_universal_2009} indicates that fundamental cognitive capacities are at play in the process. With the rise of music digitization and affective computing technologies, there has been an increasing interest in extracting emotional information from music for research, therapy \shortcite{dingle_influence_2015, bernatzky_emotional_2011} or music recommendation \shortcite{deng_emotional_2015, park_music_2015} purposes. This research falls under the term Music Emotion Recognition or MER \shortcite{chen_music_2015}. Nowadays, even music streaming platforms like Spotify come with an API that makes a wide range of features accessible and therefore open to the public.

But which features can actually predict the emotion of a song and how well perform Spotify's annotations? Building on existing literature presented in Section \ref{sec:related} we hypothesize \textit{that a multi-modal approach combining high-level auditory and lyrics-extracted features performs better than a uni-modal one} \shortcite{yang_toward_2008, hu_when_2010, hu_improving_2010}. We introduce our MER model in Section \ref{sec:experimental} before presenting and discussing the results of our exploratory and regression experiments in Sections \ref{sec:results} and \ref{sec:discussion}.

\section{Related Work}\label{sec:related}
The work on automatic information or prediction methods from music began with the rise of digital music. The annual Music Information Research Evaluation eXchange (MIREX)\footnote{A “community-based formal evaluation framework” \shortcite{downie_music_2006}} included audio music mood as a classification task for the first time in 2007 \shortcite{mirex_2007main_2007} already finding that ground-truth labeling and human mood judgment play a critical role on this task \shortcite{hu_2007_2008}. The used range of features already varied from spectral, so-called low-level features (raw frequency signal, for example, spectral centroids and SD \shortcite{yang_review_2018}) to higher-level features like “danceability” \shortcite{hu_2007_2008}, rhythm, or timbre \shortcite{yang_review_2018} that can be extracted from low-level features and are being described as closer to human perception \shortcite{panda_how_2021}. From there on, systems were consequently further being improved \shortcite{kim_music_2010}.

While the literature broadly distinguishes between three tasks of MER, i) classification into emotion categories, ii) emotion as a regression problem, and iii) variation detection throughout a song \shortcite{yang_machine_2012}, we will focus on the second problem of MER as regression since it is seen as more accurate in regards to the represented emotion \shortcite{guan_music_2012}. For this, \shortciteA{russell_circumplex_1980}’s two-dimensional model of emotion as a numerical value of valence (negative to positive) and arousal (passive to active) is the most widely used metric \shortcite{huang_bi-modal_2016, trigeorgis_adieu_2016, wang_music_2011}. For example, anger is high arousal and low valence emotion. This framework is preferred for its simplicity and is used in the present study, although it does not capture the full range of emotions \shortcite{cowen_what_2020}, which is why other MER studies \shortcite{hu_improving_2010, hu_lyric_2009} use a larger number of affective categories.

\subsection{MER as Regression Task}\label{sec:regression}

The first approaches for this regression problem were done by \shortciteA{yang_regression_2008} where they reached a valence-arousal (VA) $R^{2}$ score of up to 0.282 and 0.538 with low-level auditory features using a Support Vector Machine Regressor (SVR), similar to the SVR approach presented in \shortciteA{soleymani_emotional_2014} that reached a VA score of 0.42 and 0.52 based on supra-segmental features. Later, \shortciteA{bai_dimensional_2016}'s Random Forest Regressor (RFR) reached a VA score of 0.293 and 0.625 on 548 dimensions of low- and higher-level audio features. Other approaches include a Multiple Linear Regression (MLR) \shortcite{yang_review_2018, vatolkin_evaluation_2019}, or a Multilayer Perceptron (MLP) \shortcite{soleymani_emotional_2014, bhattarai_automatic_2019}.

The rise of fast and efficient deep learning (DL) models across a wide range of applications also impacted MER. One of the first DL bi-modal classification models was presented in \cite{jeon_music_2017} using an end-to-end approach for lyrics and raw audio without any feature engineering, fusioning the modalities mid-level. Their model outperforms uni-modal approaches at that time with an accuracy of 80\%. They also find that lyrics predict the emotion more precisely than audio. A study by \shortciteA{delbouys_music_2018}, investigates the performance of a classical feature engineering approach vs. DL. On the Deezer Mood Detection Dataset (DMDD), they extract low-level auditory features and word embeddings from a list of VA-related words and compare the mid-level (second layer or later) and late-level (weighted outputs) fusion of both of these feature classes against an end-to-end DL approach. They can show that i) mid-level fusion on DL has comparable results to the feature engineering and that ii) the multi-modal mid-level fusion outperforms both uni-modal models (audio vs. textual).

\subsection{Lyrics as Prediction Metric}
\shortciteA{yang_disambiguating_2004}, who are often seen as the first to combine audio and lyrics in MER modeling, use a Bag-of-Words (BoW) approach combined with the General Inquirer psychological lexical annotation to model lyrics. 

\begin{figure*}
    \centering
    \includegraphics[width=15cm]{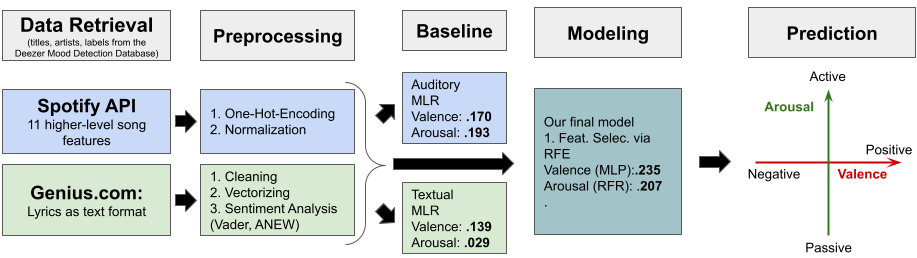}
    \caption{Pipeline of our model. Scores indicate $R^2$ on the test data.}
    \label{fig:pipeline}
\end{figure*}

\shortciteA{hu_lyric_2009, hu_improving_2010, hu_when_2010, hu_framework_2017, laurier_multimodal_2008} expanded lyrics features to include n-gram counts, stylistic features, and ANEW sentiment annotation \shortcite{bradley_affective_1998} and found that lyrics tend to outperform auditory features on emotion classification and regression tasks, though the best results were achieved by combining the features. As mentioned above, the same pattern holds in \citeA{delbouys_music_2018}'s DL model on word embeddings.

In their recent, \citeA{han_survey_2022} report that BoW, TF-IDF, and word embeddings are the most used text representation methods in MER.

\subsection{Higher-Level Features}
For the relation of features to the two predicted categories, \shortciteA{gabrielsson_influence_2001} and \shortciteA{juthi_music_2020} describe arousal as being related to tempo (fast/slow), pitch (high/low), loudness level (high/low), and timbre (bright/soft) as well as valence being related to mode (major/minor) and harmony (consonant/dissonant). \shortciteA{vatolkin_evaluation_2019}, however, state the importance of also rhythm for the prediction of valence.

\shortciteA{panda_how_2021} were the first ones to assess the correlation of 12 high-level Spotify features on the prediction accuracy of valence and arousal. They found that only three (energy, valence and acousticness) were relevant.

\subsection{Spotify}
Spotify, with its 422 million active monthly users is the largest music streaming service on the market \cite{ruby_spotify_2022}. It provides easy access to 70+ million songs \cite{ruby_spotify_2022} as well as an API that makes it possible to extract a customized dataset of annotated musical pieces\footnote{\tiny \url{https://developer.spotify.com/documentation/web-api/}}. The annotation procedure of such is unknown to the public as this is one of Spotify’s major revenue streams \shortcite{hucker_uncovering_2021}. However, the annotation technology seems automated was developed by \textit{The Echo Nest} (acquired by Spotify in 2014) based on expert-annotated data. \cite{noauthor_echo_2013}.

\section{Experimental Setup}\label{sec:experimental}
\subsection{Data} \label{ssec:data}
The song dataset used in the present study is the Deezer Mood Detection Dataset (DMDD) \shortcite{delbouys_music_2018} which holds VA scores for 18,644 songs and is based on the Million Song Dataset \cite{bertin-mahieux_million_2011} as well as tags from LastFM\footnote{\tiny \url{https://www.last.fm/}} that are related to mood \cite{hu_when_2010}. The VA scores were obtained by applying an extended ANEW (XANEW) dataset with 14,000 words and their respective VA scores \shortcite{warriner_norms_2013, bradley_affective_1998} to the tags from LastFM.

To represent the auditory modality, we retrieved the auditory features that \citeA{yang_review_2018} describe as high-level features for all available songs from the DMDD via the Spotify API\footnote{\tiny \url{https://developer.spotify.com/console/}}, resulting in 13,445 songs with 11 auditory features each. Of these, the features acousticness, danceability, energy, instrumentalness, liveness, loudness, speechiness, tempo and valence are continuous, while mode is binary, and key is categorical. We created dummy variables for key resulting in 23 auditory features overall. A detailed explanation of the features can be found in Appendix \ref{appendix:desc}. Spotify's valence annotation is derived differently from our ground-truth valence, avoiding circularity, and is also used as a predictive feature for emotion in \citeA{panda_how_2021}.  

Lastly, for the textual modality, we scraped the song lyrics for the available songs from genius.com as suggested by \citeA{li_prediction_2018}, further reducing the dataset to a final number of 12,471 songs, of which we used ca. 60\% for training, 20\% for validation and 20\% for testing.

\begin{figure*}
    \centering
    \includegraphics[width=12cm]{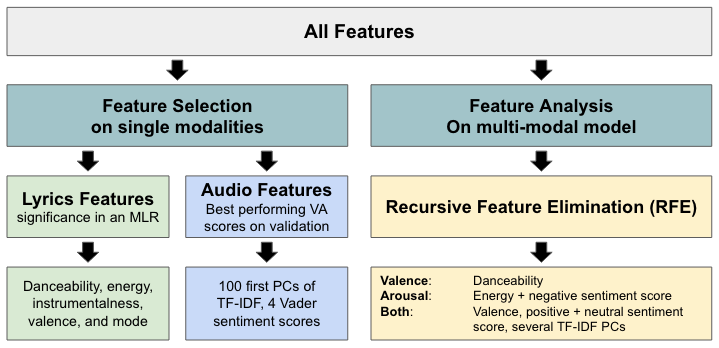}
    \caption{Feature Selection and Analysis Process}
    \label{fig:features}
\end{figure*}

\subsubsection{Extracting Lyrics Features}

To represent the lyrical information, we created three types of features. 

\begin{enumerate}
    \item \textbf{Sentiment information} consisting of positive, negative, neutral and compound scores was obtained with Vader sentiment analysis \shortcite{hutto_vader_2014}. %
    
    \item \textbf{TF-IDF features}, which are widely used \shortcite{han_survey_2022} and aim to capture the general lexical features of the song texts relating to both valence and arousal. TF-IDF unigram counts were obtained after lemmatizing the lyrics texts and reduced to 100 principal components using PCA.  
    
    \item \textbf{Extended ANEW (XANEW) features} aimed to capture the lexical basis of valence and arousal. This data set was introduced in \ref{ssec:data}. We generated two count vectors for each pre-processed lyric text and multiplied the counts by the respective VA scores. These count features were also reduced with PCA to 100 dimensions. These features were not included in the final models (see \ref{sec:featureselection}).
    
\end{enumerate}

\subsection{Feature Selection}\label{sec:featureselection}
Based on these features, we selected feature subsets for the two different modalities (see Figure \ref{fig:features}). Firstly, we tested for significance in an MLR (similar to our baseline, see Section \ref{sec:baseline}) to get the optimal combination among the auditory features resulting in the five significant predictors danceability, energy, instrumentalness, valence, and mode. 

To select lyrics features, we calculated validation VA scores for all the combinations of Vader sentiment, XANEW and TF-IDF type features (see Appendix \ref{appendix:lyrfeat}). The combination of Vader and the first 100 PCs of the TF-IDF features had the overall best performance across regression models and was used in our final modelling.

Following \shortciteA{hu_improving_2010, hu_when_2010} and especially \shortciteA{panda_how_2021} for comparability and with the aim of leading to a true multi-modal feature set, we combined all data at feature level resulting in a data matrix $M \in \mathbb{R}^{N x d}$ with $N = 12,471$ and $d = 23 + 4 + 100 = 127$.

\subsection{Models}

Based on the literature presented in Section \ref{sec:related}, we compare different regression models to evaluate the performance of uni- and multi-modal approaches for MER on high-level auditory features and lyrics. The regression learning algorithms used are all from the \texttt{scikit-learn} package \cite{pedregosa_scikit-learn_2011}: \texttt{LinearRegression}, \texttt{RandomForestRegression}, \texttt{SupportVectorRegression}, \texttt{MultilayerPerceptronRegressor}, all trained on the optimal parameters we obtained over \texttt{GridSearchCV}.

\subsubsection{Baseline}\label{sec:baseline}
We created a Multiple Linear Regression (MLR) baseline for both uni- and multi-modal prediction of VA scores. 
For the audio baseline, we firstly we used all auditory features as input. Secondly, we performed the MLR on only statistically significant ($p<0.05$) coefficients. Then for the lyrics baseline, we performed the MLR on the selected lyrics features (sentiment and TF-IDF). Lastly, we combined the significant audio features and the selected lyrics features in MLR to form the multi-modal baseline. 
\subsection{Post-Hoc Feature Analysis}\label{sec:analysis}
Additionally, for investigating the set of most significant features overall, we applied Recursive Feature Elimination (RFE) \shortcite{granitto_recursive_2006} using the \texttt{sklearn.feature\_selection.RFE} package \shortcite{pedregosa_scikit-learn_2011} on an MLR model. Furthermore, we created a MLR model with all audio features + compound sentiment scores included, for the purpose of analyzing coefficients (see Table \ref{tab:coef}).

\section{Results}\label{sec:results}
We evaluate the predictive power of different regression models in terms of $R^{2}$ as applied in the aforementioned literature (e.g. \shortciteA{hu_2007_2008, hu_lyric_2009, kim_music_2010}).

\subsection{Model Results}

Table \ref{tab:results_modalities} provides a summary of test results for our regression models with different modalities. The highest scores overall are achieved with multi-modal features for both valence and arousal. This is more pronounced for valence: multi-modality significantly improves $R^{2}$ compared to both audio- and lyrics-only features. Within the uni-modal models however, audio features are slightly better at predicting valence than lyrics features.

\begin{table}[htb!]
\begin{tabular}{|l|l||l|l|}
\hline
\textbf{Mode} & \textbf{Model} & \textbf{Valence} & \textbf{Arousal} \\ \hline \hline
\multirow{4}{*}{Audio} & MLR & 0.170 & 0.193 \\ \cline{2-4} 
 & RFR & 0.171 & \textbf{0.204} \\ \cline{2-4} 
 & SVR & 0.165 & 0.203 \\ \cline{2-4} 
 & MLP & \textbf{0.176} & 0.203 \\ \hline
\multirow{4}{*}{Lyrics} & MLR & \textbf{0.139} & \textbf{0.029} \\ \cline{2-4} 
 & RFR & 0.121 & 0.027 \\ \cline{2-4} 
 & SVR & 0.042 & -0.074 \\ \cline{2-4} 
 & MLP & 0.117 & 0.020 \\ \hline
\multirow{4}{*}{Multi-modal} & MLR & \textbf{0.236} & 0.190 \\ \cline{2-4} 
 & RFR & 0.224 & \textbf{0.207} \\ \cline{2-4} 
 & SVR & 0.208 & 0.154 \\ \cline{2-4} 
 & MLP & 0.235 & 0.196 \\ \hline
\end{tabular}
\caption{$R^2$ test scores for all uni- and multi-modal models based on selected feature subsets as described in \ref{sssec:num1}.}

\label{tab:results_modalities}
\end{table}

The arousal results are more complicated. Although the highest arousal score is achieved with multi-modal RFR, all the other regression algorithms perform better at arousal prediction with uni-modal (audio) features. This suggests that adding lyrics features may interfere and compound error for arousal, because we also see that the lyrics-only arousal scores are practically 0. 

No single regression algorithm emerged as the best, although SVR consistently performed worst. This might be due to insufficient hyperparameter optimization, as testing a larger number of parameter combinations was very time-consuming for this type of model. 

Surprisingly, for lyrics-only valence and arousal as well as multi-modal valence, our more complex regression algorithms did not outperform the baseline MLR.

\subsection{Feature Analysis}
Based on RFE, the most important features were valence, the positive and neutral sentiment score and several PCs from the TF-IDF analysis for both target variables (see Figure \ref{fig:features}) as well as danceability for valence and energy and the negative sentiment score for arousal. 

Considering the p-values of coefficients in MLR, valence has 7 significant predictors, where arousal has 6 as can be seen in Table \ref{tab:coef}.

\begin{table}[htb!]
\begin{tabular}{|l||l|l|}
\hline
\textbf{Feature} & \textbf{Valence} & \textbf{Arousal} \\ \hline \hline
Constant & \textbf{-1.6885*} & \textbf{-0.9836*} \\ \hline
Danceability & \textbf{0.6915*} & \textbf{-0.3266*} \\ \hline
Energy & \textbf{0.6378*} & \textbf{1.4254*} \\ \hline
Loudness & -0.0091 & -0.0073\\ \hline
Speechiness & -0.1101 & \textbf{0.3952*} \\ \hline
Acousticness & \textbf{0.1649*} & 0.0207 \\ \hline
Instrumentalness & 0.0929 & \textbf{-0.3278*} \\ \hline
Liveness & \textbf{0.1916*} & 0.0207 \\ \hline
Valence & \textbf{1.0901*} & \textbf{0.5158*} \\ \hline
Tempo & 0.0005 & 0.0004 \\ \hline
Mode & \textbf{0.0977*} & \textbf{0.1272*} \\ \hline
Compound sentiment & \textbf{0.2275*} & -0.0051 \\ \hline
\end{tabular}
\caption{Coefficients for MLR. *significant with p \textless 0.05}
\label{tab:coef}
\end{table}

\subsubsection{Selected Features Performance} \label{sssec:num1}

Table \ref{tab:features} compares VA scores of the MLP with all features vs. selected features for each modality. The feature subsets are listed here:  

\begin{enumerate}
    \item all\_features$_{A}$ = \{Acousticness, Danceability, Energy, Instrumentalness, Key, Liveness, Loudness, Mode, Speechiness, Tempo, Valence\}
    \item selected$_{A}$ = \{Danceability, Energy, Instrumentalness, Valence, Mode\}
    \item all\_features$_{L}$ = \{200 PCs ANEW scores, 100 PCs TF-IDF, 4 Vader sentiment scores\}
    \item selected$_{L}$ = \{100 PCs of TF-IDF, 4 Vader sentiment scores\}
\end{enumerate}

\noindent We chose to illustrate these results using the MLP, as it has a consistent  performance and takes little time to optimize and train. We see  that using selected features gives a small but consistent boost to the $R^{2}$ for all modalities, so we trained our final models using these features.

Overall, we can see from Table \ref{tab:features} that all subsets of selected features perform always better in comparison to all the features.

\begin{table}[!htb]
\begin{tabular}{|p{1cm}||p{2.3cm}||l|l|}
\hline
                              & \textbf{Feature set} & \textbf{Valence}                       & \textbf{Arousal} \\ \hline \hline
                              & all\_features$_{A}$ & 0.163                        & 0.193 \\ \cline{2-4} 
\multirow{-2}{*}{Audio}       & selected$_{A}$ & \textbf{0.176}               & \textbf{0.203} \\ \hline
                              & all\_features$_{L}$ & 0.091                       & 0.009 \\ \cline{2-4} 
\multirow{-2}{*}{Lyrics} & selected$_{L}$ & \textbf{0.117} & 0.019 \\ \hline
                              & all\_features$_{A}$ + all\_features$_{L}$ & \cellcolor[HTML]{FFFFFF}0.230 & 0.193           \\ \cline{2-4} 
\multirow{-2}{*}{Multi} & selected$_{A}$ + selected$_{L}$ & \textbf{0.235} & \textbf{0.196} \\ \hline
\end{tabular}
\caption{Comparison of MLP $R^{2}$-scores for different feature subsets}
\label{tab:features}
\end{table}

\section{Discussion}\label{sec:discussion}
Our main results can be briefly summarized in two notions. One, both uni-modal lyrics features and uni-modal audio features reasonably predict valence, although a multi-modal approach outperforms either modality individually. And two, predicting arousal is hard to do with lyrics features, since audio features alone perform almost as well as the multi-modal approach. 

\subsection{Effect of Features}

\textbf{Valence.} As seen in Table \ref{tab:coef}, danceability, energy, valence (Spotify), mode, acousticness, liveness  and  compound sentiment positively contribute to the valence score, while the first three are by far the strongest predictors.

\textbf{Arousal.} The features energy, speechiness, mode, and valence positively contribute, while danceability and instrumentalness have negative coefficients. Energy is by far the strongest predictor.

Both of these findings are consistent with \citeA{panda_how_2021} in regards to energy, valence and acousticness. However, unlike them, we found that  danceability, instrumentalness, mode and speechiness are also relevant for MER. Especially the difference in findings about the importance of danceability is interesting. Where in our baseline model danceablility has one of the highest coefficients, \citeA{panda_how_2021} found that danceability has the lowest feature weight.

Danceability captures tempo, rhythm and beat aspects of a song, where perceptual features such as perceived loudness and timbre contribute to energy according to Spotify's descriptions (see appendix \ref{appendix:desc}). Danceability being a strong predictor of valence supports previous findings about the importance of rhythmic properties for predicting valence \cite{yang_regression_2008,vatolkin_evaluation_2019}. Our finding that energy is a strong predictor of arousal indirectly supports research stating timbre is related to arousal \cite{gabrielsson_influence_2001,yang_regression_2008}. As energy also captures perceived loudness, the association between loudness and arousal \cite{gabrielsson_influence_2001,yang_regression_2008} is also supported by our results.

The finding that Spotify’s valence is the strongest predictor of valence is an indication that it is indeed somehow measured by the same underlying construct, despite being constructed in different ways. Furthermore, we find a positive coefficient of mode and predicted valence, which corresponds to traditional associations of major keys being related to positive emotions and minor keys to negative ones \cite{gabrielsson_influence_2001,panda_audio_2020}.

The positive coefficient of sentiment score for predicting valence could be interpreted as that songs with more positive lyrics have higher valence scores and songs with more negative lyrics have more lower valence scores.

Surprisingly, XANEW features were among the least predictive features based on RFE and VA validation scores, which is why they were removed from the final models. The ground-truth VA scores are based on tag analysis with XANEW \shortcite{delbouys_music_2018}, and studies \citeA{hu_improving_2010} yielded good results with ANEW features. The main difference between XANEW features and VA scores is that the former is based on tags from LastFM, where the latter is based on the lyrics. This suggests that words in lyrics of the songs substantially differ from the tags people give a song on LastFM. TF-IDF may have been more successful in finding distinguishing words because it takes into account the entire lyric vocabulary.

\subsection{Multi-modal vs. Uni-modal}

\textbf{Valence.} Our results for modality are generally consistent with the literature (\shortcite{yang_toward_2008, hu_improving_2010, delbouys_music_2018}), where multi-modal MER models outperform uni-modal ones. However, in our case this is only unambiguously true for valence, where lyrics and audio features appear to have complementary effects when combined. The fact that audio seems to contribute more to valence (based on uni-modal scores) is consistent with \shortciteA{ali_songs_2006}, \shortciteA{delbouys_music_2018} and \shortciteA{xue_multimodal_2015}, as well as anecdotal evidence \cite{heinrich_22_2019} that people pay more attention to music than lyrics, which would influence the listener annotations that our target variables are based on. On the other hand, older studies like \shortciteA{hu_improving_2010, hu_when_2010} do report that lyrics can outperform audio for valence prediction. However they make use of more extensive feature engineering and more fine-grained emotion categories (predicted in binary classification tasks), which limits comparability. 

\textbf{Arousal.} With regards to arousal prediction, the multi-modal advantage is not clear. Only one of our regression models performs better multi-modally, while the other three perform better with only audio features. Lyrics features seem to either not contribute or actively interfere with audio features to give lower multi-modal performance. This is consistent with \shortciteA{delbouys_music_2018}'s findings that arousal is mostly determined by audio features such as rhythm and tempo. \shortciteA{malheiro_emotionally-relevant_2018} succeed in classifying songs with regards to both arousal and valence exclusively from lyrics, but with a small dataset and very extensive feature engineering, especially including more stylistic and structural features, which we pre-processed away in our BoW TF-IDF approach. On the other hand, \shortciteA{delbouys_music_2018}, who do not report pre-processing lyrics but feed them directly into a variety of neural networks, achieve comparably low arousal scores for their lyrics-only models on the same dataset. 

\subsection{Summary and Future Directions}

Lastly, we want to shed light on possible reasons for our model's performance as well as limitations and give an outlook for future research direction.

\textbf{Early Feature Fusion.} As stated in \ref{sec:featureselection}, we chose to combine our features at feature level, following \shortciteA{hu_improving_2010, hu_when_2010} and \shortciteA{panda_how_2021}. However, \shortciteA{yang_toward_2008} states that a potential difficulty of early fusion is achieving a truly common representation. Especially the combination of radically different feature types like TF-IDF vectors and auditory features could be responsible for our results. An exploratory investigation of late-fusion approaches (see e.g. \shortciteA{yang_toward_2008, delbouys_music_2018} for this dataset is left for future investigation. Another interesting direction for future research is investigating interaction effects of the features we used.

\textbf{Deep Learning as State-of-the-Art.}
Interestingly, we find that out results align very well with the findings of \shortciteA{delbouys_music_2018}'s comparision of classical feature engineering based approaches and DL models on the same dataset. We reach competetive scores on valence on the audio-only prediction with a MLP (0.176) compared to their convolutional network (0.179) while our multi-modal MLR (0.236) even outperforms their middle-fusion DL network (0.219) on valence.  
This suggests that using higher-level features makes our model more interpretable, while not compromising on performance.

\textbf{Vague Annotation Standard.}
That we obtain similar results as \shortciteA{delbouys_music_2018} indicates that part of the limitations may be related to the dataset itself. The fact that Spotify's valence doesn't correlate to the predicted valence very well ($R^{2}$ = 0.35) means it also may be hard to measure and evaluate accurately. The VA annotations obtained through LastFM tags in the \shortciteA{delbouys_music_2018} were not checked or validated with human annotation, so their quality is unknown. Furthermore, the target VA variables assume that there is a single value for one whole song rather than displaying a possible variation of valence and arousal within the piece, while \shortciteA{yang_review_2018} suggest that a song's sentiment can vary throughout the piece, which falls under the MER subdomain of music emotion variation detection (MEVD) \shortcite{han_survey_2022}.

\textbf{Audio and Predictions.} Even though we find some improvement to the overall scores when including the selected Spotify features, other approaches that directly source from the song's spectral features (e.g. \shortciteA{hu_2007_2008, hu_improving_2010, soleymani_emotional_2014}), seem to achieve better results at least on valence scores of up to 0.42 already in 2014. This leads to the conclusion in alignment with \shortciteA{panda_how_2021} that there is still some room for improvement and clarity for Spotify features and that valence and arousal are still rather hard to predict from high-level features.

\textbf{Lyrics and Predictions.}
Already the first music mood classification task, \shortciteA{mirex_2007main_2007}, found that MER depends highly on the ground-truth labeling and human judgement, including how much people pay attention to the lyrics. An example for this is "Hey-Ya" by OutKast, that sounds happy and upbeat (which corresponds to a high valence and arousal score in our dataset), whereas the artists themselves attribute a more negative meaning to it \cite{archive-corey-moss_road_2004}. Henceforth, the true, multi-modal valence of such songs is hard to determine, both for human judges as well as predictive models. Aggravating factors to this are the lyrics shortness and the use of rhetorical figures like metaphors, oxymorons, and irony that makes it even harder to capture the lyrics' true sentiment. For future direction we suggest a more advanced sentiment analysis.

\section{Conclusion}\label{sec:conclusion}
We have shown that we can combine audio and textual modalities to predict valence and arousal for a large dataset of songs, improving in particular valence prediction relative to uni-modal approaches. Predicting arousal from lyrics proved unsuccessful with the features we used and may require more extensive feature engineering or DL approaches. On average, audio-only models were better than multi-modal ones at predicting arousal based on auditory features such as energy, speechiness and mode.

\bibliographystyle{apacite}
\bibliography{refs}

\appendix
\clearpage

\onecolumn

\section{Feature description}\label{appendix:desc}

\begin{table}[!htb]
    \centering
    \begin{tabular}{|l|>{\small}p{13cm}|}
    \hline
     Acousticness & A confidence measure from 0.0 to 1.0 of whether the track is acoustic. 1.0 represents high confidence the track is acoustic. 0 $<=$ x $<=$ 1 \\ \hline
     Danceability & Describes how suitable a track is for dancing based on a combination of musical elements including tempo, rhythm stability, beat strength, and overall regularity. A value of 0.0 is least danceable and 1.0 is most danceable. \\ \hline
     Energy & Energy is a measure from 0.0 to 1.0 and represents a perceptual measure of intensity and activity. Typically, energetic tracks feel fast, loud, and noisy. For example, death metal has high energy, while a Bach prelude scores low on the scale. Perceptual features contributing to this attribute include dynamic range, perceived loudness, timbre, onset rate, and general entropy. \\ \hline
     Instrumentalness & Predicts whether a track contains no vocals. "Ooh" and "aah" sounds are treated as instrumental in this context. Rap or spoken word tracks are clearly "vocal". The closer the instrumentalness value is to 1.0, the greater likelihood the track contains no vocal content. Values above 0.5 are intended to represent instrumental tracks, but confidence is higher as the value approaches 1.0. \\ \hline
     Key & The key the track is in. Integers map to pitches using standard Pitch Class notation. E.g. 0 = C, 1 = C/Dx, 2 = D, and so on. If no key was detected, the value is -1. -1 $<=$ x $<=$ 11 \\ \hline
     Liveness  & Detects the presence of an audience in the recording. Higher liveness values represent an increased probability that the track was performed live. A value above 0.8 provides strong likelihood that the track is live. \\ \hline
     Loudness & The overall loudness of a track in decibels (dB). Loudness values are averaged across the entire track and are useful for comparing relative loudness of tracks. Loudness is the quality of a sound that is the primary psychological correlate of physical strength (amplitude). Values typically range between -60 and 0 db. \\ \hline
     Mode & Mode indicates the modality (major or minor) of a track, the type of scale from which its melodic content is derived. Major is represented by 1 and minor is 0. \\ \hline
     Speechiness & Speechiness detects the presence of spoken words in a track. The more exclusively speech-like the recording (e.g. talk show, audio book, poetry), the closer to 1.0 the attribute value. Values above 0.66 describe tracks that are probably made entirely of spoken words. Values between 0.33 and 0.66 describe tracks that may contain both music and speech, either in sections or layered, including such cases as rap music. Values below 0.33 most likely represent music and other non-speech-like tracks. \\ \hline
     Tempo & The overall estimated tempo of a track in beats per minute (BPM). In musical terminology, tempo is the speed or pace of a given piece and derives directly from the average beat duration. \\ \hline
     Valence & A measure from 0.0 to 1.0 describes the musical positiveness conveyed by a track. Tracks with high valence sound more positive (e.g. happy, cheerful, euphoric), while tracks with low valence sound more negative (e.g. sad, depressed, angry). 0 $<=$ x $<=$ 1 \\ \hline
        \end{tabular}
    
    \caption{Used features and description taken from the \href{https\:\/\/developer.spotify.com\/documentation\/web-api\/reference\/\#/operations\/get-several-audio-features}{Spotify documentation}.}

\end{table}

\section{Performance of Lyrics Features}\label{appendix:lyrfeat}

\begin{table}[!htb]
\centering
\resizebox{\textwidth}{!}{
\begin{tabular}{l|ll|ll|ll|ll}
                 & \multicolumn{2}{c}{\textbf{LinReg}} \vline &  \multicolumn{2}{c}{\textbf{RF}} \vline &  \multicolumn{2}{c}{\textbf{SVR}} \vline & \multicolumn{2}{c}{\textbf{MLP}}  \\ \hline
\textbf{Features}         & Valence           & Arousal           & Valence       & Arousal           & Valence               & Arousal               & Valence             & Arousal           \\ \hline
tfidf            & 0.0705            & 0.0188            & 0.0631        & 0.0251            & -0.0138               & -0.0099               & -0.8320             & -2.0411             \\
anew             & -0.0146           & -0.0138           & 0.0060        & -0.0091           & -0.0255               & -0.0336               & 0.0121              & -0.0026             \\
vader            & 0.0478            & -0.0025           & -0.0049       & -0.0765           & 0.0076                & -0.0003               & 0.0536              & 0.0041              \\
tfidf+anew       & -0.0028           & -0.0291           & 0.0427        & -7.0790e-06       & 0.0765                & -0.0056               & \textbf{0.0962}          & 0.0167              \\
tfidf+vader      & \textbf{0.0931 }           & \textbf{0.0159}          & \textbf{0.0842 }       & \textbf{0.0267}            & 0.0817                & \textbf{0.0044}               & 0.0921              & \textbf{0.0216}             \\
anew+vader       & 0.0238            & -0.0187           & 0.0536        & -0.0056           & 0.0284                & -0.0223               & 0.0518              & -0.0038             \\
tfidf+anew+vader & 0.0304            & -0.0251           & 0.0796        & 0.0019            & \textbf{0.0890}              & 0.0041                & 0.0911              & 0.0089             
\end{tabular}}

\caption{Validation $R^{2}$ scores for valence and arousal for different combinations of lyrics features. Tf-idf and Vader sentiment emerged as the best combination of features. The regression models were run on their default parameters.}

\end{table}

\end{document}